\documentstyle[12pt]{article}
\newcommand{\be}{\begin{equation}}
\newcommand{\ee}{\end{equation}} 
\newcommand{\eins}{\mbox{$\rule{2.5mm}{0.1mm}
                          {\hspace{-2.7mm}1}
                          {\hspace{-0.2mm}\rule{0.07mm}{2.7mm}}$}}
\newcommand{\vslash}{\mbox{$\not{\hspace{-0.8mm}v}$}}          
 
\begin{document}
\bibliographystyle{references}
\baselineskip 18pt
\pagenumbering{arabic}         
\begin{titlepage}
\begin{flushright}
 IC/96/35\\
 MZ-TH/96-10
\end{flushright}

\begin{center}

  {\large \bf One--pion transitions between heavy baryons in the
constituent quark model}   

 \vspace*{0.7cm}

 {\large
  F. Hussain}

 {\large
  International Center for Theoretical Physics, Trieste, Italy}\\   

  \vspace*{0.3cm}

{\large J.G. K\"{o}rner}\\

{\large
  Institut f\"{u}r Physik, Johannes Gutenberg-Universit\"{a}t\\
  Staudinger Weg 7, D-55099 Mainz, Germany}\\
 \vspace*{0.4cm}
and\\
\vspace*{0.3cm}    

{\large Salam Tawfiq}

 {\large
  International Center for Theoretical Physics, Trieste, Italy}\\   

\vspace*{0.4cm}    
{\large July 1999}
      
 \vspace*{1cm}

\end{center}
 
\begin {abstract}
Single pion transitions of
{\it S} wave to {\it S} wave, {\it P} wave to {\it S} wave and {\it P} wave
to {\it P} wave
heavy baryons
are
analyzed in the framework of the Heavy Quark Symmetry limit (HQS). We then
use a constituent quark model picture for the light diquark system
with an underlying $SU(2N_{f}) \otimes O(3)$ symmetry to reduce the number of
the HQS coupling factors required to describe these transitions.
A single constituent quark model $p$-wave coupling
is necessary to describe transitions among the 
{\it S} wave ground states. One $s$-wave and one $d$-wave coupling 
factors are required to determine each of the transitions from the orbitally 
symmetric K-multiplet and from the orbitally antisymmetric k-multiplet down 
to the ground state. {\it P} wave to {\it P} wave single pion transitions
are described
by altogether eight constituent quark model coupling constants. 
We also estimate
 decay rates of some single pion transitions between charm baryon states.
 \end{abstract}
 
\end{titlepage}

\newpage
                              
\section{Introduction}
\begin{table}
\begin{center}
\caption{\label{tab1}Spin wave functions (s.w.f.) and flavor wave functions 
  (f.w.f.) of heavy $\Lambda$-type and $\Sigma$-type $S$- and $P$-wave heavy 
  baryons.}
\vspace{5mm}
\renewcommand{\baselinestretch}{1.2}
\small \normalsize
\begin{tabular}{|c|ccccc|}\hline\hline
&\begin{tabular}{c}
    light diquark s.w.f.\\
    ${\hat\phi}^{\mu_1\cdots\mu_j}$
  \end{tabular}
&\begin{tabular}{c}
    f.w.f.\\
    $T$
  \end{tabular}
&$j^P$
&\begin{tabular}{c}
    heavy diquark s.w.f.\\
    $\psi_{\mu_1\cdots\mu_j}$
  \end{tabular}
&$J^P$\\\hline\hline
\multicolumn{6}{|l|}{$S$-wave states ($l_k=0,l_K=0$)}\\
$\Lambda_Q$&$\chi^0$&$T^{(3^*)}$&$0^+$&$u$&$\frac12^+$\\\hline
$\Sigma_Q$&$\chi^{1,\mu}$&$T^{(6)}$&$1^+$
 &$\begin{array}{r}
     \frac1{\sqrt3}\gamma^\perp_\mu\gamma_5u\\
     u_\mu
  \end{array}$
 &$\begin{array}{c}
     \frac12^+\\
     \frac32^+
     \end{array}$\\\hline\hline
\multicolumn{6}{|l|}{$P$-wave states ($l_k=0,l_K=1$)}\\
$\Lambda_{QK1}$&$\chi^0K_\perp^\mu$&$T^{(3^*)}$&$1^-$
  &$\begin{array}{r}
    \frac1{\sqrt3}\gamma^\perp_\mu\gamma_5u\\
    u_\mu
  \end{array} $
  &$\begin{array}{c}
    \frac12^-\\
    \frac32^-
  \end{array}$\\\hline
$\Sigma_{QK0}$&$\frac1{\sqrt3}\chi^1\cdot K_\perp$&$T^{(6)}$&$0^-$
&$u$&$\frac12^-$\\\hline
$\Sigma_{QK1}$&$\frac i{\sqrt2}\varepsilon(\mu\chi^1K_\perp v)$&$T^{(6)}$
&$1^-$
  &$\begin{array}{r}
    \frac1{\sqrt3}\gamma^\perp_\mu\gamma_5u\\
    u_\mu
  \end{array}$
  &$\begin{array}{c}
    \frac12^-\\
    \frac32^-
  \end{array}$\\\hline
$\Sigma_{QK2}$&$\frac12\{\chi^{1,\mu_1}K_\perp^{\mu_2}\}_0$&$T^{(6)}$
&$2^-$
  &$\begin{array}{r}
    \frac1{\sqrt{10}}\gamma_5\gamma^\perp_{\{\mu_1}
    u_{\mu_2\}_0}^{\phantom{\perp}}\\
    u_{\mu_1\mu_2}
  \end{array}$
 &$\begin{array}{c}
   \frac32^-\\
   \frac52^- 
 \end{array}$\\\hline\hline
\multicolumn{6}{|l|}{$P$-wave states ($l_k=1,l_K=0$)}\\
$\Sigma_{Qk1}$&$\chi^0k^\mu_\perp$&$T^{(6)}$&$1^-$
  &$\begin{array}{r}
    \frac1{\sqrt3}\gamma^\perp_\mu\gamma_5u\\
    u_\mu
  \end{array}$
  &$\begin{array}{c}
    \frac12^-\\
    \frac32^-
  \end{array}$\\\hline
$\Lambda_{Qk0}$&$\frac1{\sqrt3}\chi^1\cdot k_\perp$&$T^{(3^*)}$&$0^-$
&$u$&$\frac12^-$\\\hline
$\Lambda_{Qk1}$&$\frac i{\sqrt2}\varepsilon(\mu\chi^1k_\perp v)$&$T^{(3^*)}$
&$1^-$
  &$\begin{array}{r}
    \frac1{\sqrt3}\gamma^\perp_\mu\gamma_5u\\
    u_\mu
  \end{array}$
  &$\begin{array}{c}
    \frac12^-\\
    \frac32^-
    \end{array}$\\\hline
$\Lambda_{Qk2}$&$\frac12\{\chi^{1,\mu_1}k_\perp^{\mu_2}\}_0$&$T^{(3^*)}$
&$2^-$
  &$\begin{array}{r}
    \frac1{\sqrt{10}}\gamma_5\gamma_{\{\mu_1}^\perp
      u_{\mu_2\}_0}^{\phantom{\perp}}\\
    u_{\mu_1\mu_2}
  \end{array}$
  &$\begin{array}{c}
    \frac32^-\\
    \frac52^-
  \end{array}$\\\hline\hline
\end{tabular}
\renewcommand{\baselinestretch}{1}
\small \normalsize
\end{center}
\end{table}            
Based on Heavy Quark Symmetry (HQS) and the $SU(2N_f)\times O(3)$ light
diquark symmetry, the construction of heavy baryon wave functions in the limit
$m_Q\rightarrow \infty$, was formulated in \cite{htk,kkp}.
The analysis of the current-induced {\it S} wave bottom baryon to both S
wave and
{\it P} wave charm baryon transitions, in the constituent quark model, was
reported in
\cite{hklt}. We follow the same procedure to study the single pion transitions
between heavy charm or bottom baryons. The physics of the single--pion
transitions between heavy baryons is quite simple: the pion is emitted from the
light diquark while the heavy
quark propagates unaffected by the pion emission process. Since the heavy
baryon is infinitely massive it will not recoil when emitting the 
pion, i.e. the
 velocity of the heavy quark and, thereby, that of the heavy baryon remains
unchanged.

 Heavy Quark Symmetry allows us to write the heavy baryon spin wave function
  for arbitrary orbital angular momentum as \cite{htk,kkp,hklt} 
\be
\Psi_{\alpha\beta\gamma}(v,k,K)=
\phi^{\mu_{1}\cdots\mu_{j}}_{\alpha\beta}(v,k,K)\psi_{\mu_{1}\cdots\mu_{j};\gamma}\,,
\ee
where we have neglected flavor factors for the moment. The ``superfield"
heavy-side baryon spin wave function
$\psi_{\mu_{1}\cdots\mu_{j};\gamma}$ stands for the two spin wave
functions corresponding to baryon spin $\{j-1/2,j+1/2\}$, where $j$ is the 
total angular momentum of the light diquark system. The Dirac indices 
$\alpha$, $\beta$ and $\gamma$ refer to the two light quarks and the
heavy quark, respectively, and the $\mu_{i}$'s are Lorentz indices. Here, $v$
is the velocity of the baryon, while $k=\frac{1}{2}(p_{1}-p_{2})$ and
$K=\frac{1}{2}(p_{1}+p_{2}-2p_{3})$ are the two independent relative
momenta which can be formed from the two light quark momenta $p_{1}$ and
$p_{2}$ and from the heavy quark momentum $p_{3}$. Furthermore, the
light diquark wave function can be written as
\be
\phi^{\mu_{1}\cdots\mu_{j}}_{\alpha\beta}(v,k,K)={\hat
\phi}^{\mu_{1}\cdots\mu_{j}}_{\delta\rho}(v,k,K)A^{\delta\rho}_{\alpha\beta}\,,
\ee
where the ${\hat\phi}^{\mu_{1}\cdots\mu_{j}}_{\delta\rho}(v,k,K)$ are
spin projection operators which project out particular spin and parity
states of the diquark from the unknown orbital wave functions $A$.
In the following, we shall refer to the
${\hat\phi}^{\mu_{1}\cdots\mu_{j}}_{\delta\rho}(v,k,K)$ as the
light diquark spin wave functions (s.w.f.). The spin wave
functions for both the heavy-side and light diquark system, for {\it S} 
wave and 
{\it P} wave baryons, are listed in Table \ref{tab1}. In this table,
${\chi}^0 =\frac{1}{2\sqrt{2}}[(\vslash+1) \gamma_5C]$ and
${\chi}^{1,\mu}=\frac{1}{2\sqrt{2}}
[(\vslash +1) \gamma_\perp^\mu C]$ with $C$ being the charge
conjugation operator. Details of the normalization of the light diquark
and heavy-side spin wave functions can be found in \cite{kkp} and
\cite{hklt}.

In Table \ref{tab1}, we have included the symmetric and antisymmetric
light diquark flavor wave functions $T^{(6)}_{ab}$ and 
$T^{(3^{*})}_{ab}$ respectively. They transform as
the sextet
($\Sigma$-type) and anti-triplet
($\Lambda$-type) representation of flavor $SU(3)$.
Explicit representations of these functions, in terms of their 
quark content, will be given later on.   
The ${\hat\phi}^{\mu_{1}\cdots\mu_{j}}_{\alpha\beta}(v,k,K)\otimes T_{ab}$
are constructed ensuring overall symmetry with respect to
$colour\otimes flavour\otimes
spin\otimes orbital$. It is not difficult to see from
Table \ref{tab1} that, including the flavor factors and
defining the indices $(A=\alpha,a\,;\,B=\beta,b)$, the 
${\hat\phi}^{\mu_{1}\cdots\mu_{j}}_{AB}=
{\hat\phi}^{\mu_{1}\cdots\mu_{j}}_{\alpha\beta}T_{ab}$ are symmetric for
{\it S} wave states under interchange of the indices $A$ and $B$ whereas
they are
symmetric (antisymmetric) for {\it P} wave states depending on whether they
are
functions of $K$ ($k$). Since $\alpha$ and $\beta$ are essentially two
component indices in $\chi^{0}$ and $\chi^{1,\mu}$ because of the
Bargmann-Wigner equations \cite{htk,hklt},
the light diquark spin-flavor wave functions
${\hat\phi}^{\mu_{1}\cdots\mu_{j}}_{AB}$ transform as irreducible
representations of $SU(6)\otimes O(3)$ if $N_{f}=3$ is the number of
light flavors. The three different blocks in Table \ref{tab1} correspond to
the three $SU(6)\otimes O(3)$ irreducible representations
${\bf 21}\otimes {\bf 1}$, ${\bf 21}\otimes {\bf 3}$ and ${\bf 15}\otimes
{\bf 3}$ respectively.
This will be an important fact to remember later.

In the next section, we shall present the HQS predictions for single pion 
transitions among ground states and from {\it P} wave to {\it S} wave
states.
Sec. 3
is devoted to derive the constituent quark model relations using the 
$SU(2N_{f}) \otimes O(3)$ symmetry of the light degrees of freedom. Various 
phenomenological predictions for charmed baryons strong decays are presented 
in Sec. 4. 
Section 5 contains some concluding remarks. Moreover, the constituent 
quark model is also employed to reduce the number of heavy quark symmetry 
coupling constants for {\it P} wave to {\it P} wave transitions in Appendix
A. 
Finally, we use the quantum theory of angular 
momentum in Appendix B to rewrite our constituent quark model results
on the one-pion transitions in terms of a general master formula
containing a product of 6j- and 9j-symbols. 
\section{Heavy Quark Symmetry Relations}
In the heavy quark limit, the orbital momenta of the pion
relative to the diquark $l_{\pi}$ and relative to the baryon $L_{\pi}$ are
identical ( $l_{\pi}=L_{\pi}$) .                                 
Using Heavy Quark Symmetry the one--pion transition amplitudes between heavy baryons can then be
written as \cite{kkp}
\begin{table}[h]
\caption{\label{tabtens}Tensor structure of covariant pion couplings for the
allowed diquark transitions. The fourth column gives the values of the rate 
factors $ \mid c(j_1,j_2,l)\mid^2 $ in the rate formula Eq. (\ref{rate-gen}).}
\vspace{5mm}
\renewcommand{\baselinestretch}{1.2}
\small \normalsize
\begin{center}
\begin{tabular}{||c|c|c|c||}
\hline \hline
     Orbital Wave & Diquark Transition & Covariant Coupling & Rate Factor \\
  $  l_{\pi} $ & $j_1^{p_1}\rightarrow j_2^{p_2}+\pi$&$ t^{l_{\pi}}_{\mu_1 \dots \mu_{j_1};
  \nu_1 \dots \nu_{j_2}} $ & $ \mid c(j_1,j_2,l)\mid^2 $ \\
 \hline \hline
  s-wave & $ 0^- \rightarrow 0^+ $  & scalar & $ 1/2 $\\
  $ (l_{\pi}=0) $ & $ 1^- \rightarrow 1^+ $ & $ g_{\perp \mu_1\nu_1} $ & $ 1/2 $\\
 \hline
  & $ 1^{\pm} \rightarrow 0^{\pm} $ & $ p_{\perp \mu_1} $ & $ 1/6 $\\
  p-wave  & $ 1^{\pm} \rightarrow 1^{\pm} $&$ i\varepsilon (\mu_1\nu_1 p
v)$ & $ 1/3 $ \\
  $ (l_{\pi}=1) $  & $ 2^{-} \rightarrow 1^{-} $ & $ g_{\perp
\mu_1\nu_1}p_{\perp \mu_2} $ & $ 1/6 $  \\
  & $ 2^{-} \rightarrow 2^{-} $ & $ i\varepsilon (\mu_1\nu_1 p v)g_{\perp
\mu_2\nu_2} $ & $ 1/4 $ \\
 \hline
 d-wave & $ 1^{-} \rightarrow 1^{+} $ & $ T_{\mu_1\nu_1} $ & $ 1/9 $  \\
 $(l_{\pi}=2) $ & $ 2^{-} \rightarrow 0^{+} $ & $ T_{\mu_1\mu_2} $ & $ 1/15 
$ \\
    & $ 2^{-} \rightarrow 1^{+} $ & $ i\varepsilon (\mu_1\nu_1\rho\sigma)v^{\sigma}
    T^{\rho}_{\mu_2} $ & $ 1/10$ \\
\hline
f-wave & $ 2^{-} \rightarrow 1^{-} $ & $ T_{\mu_1\mu_2\nu_1} $ & $ 1/25$ \\
$(l_{\pi}=3) $ & $ 2^{-} \rightarrow 2^{-} $ & $ i\varepsilon (\mu_1\nu_1\rho
\sigma)v^{\sigma}T^{\rho}_{\mu_2\nu_2} $ & $ 1/15 $  \\

\hline \hline
\end{tabular}
\renewcommand{\baselinestretch}{1}
\small \normalsize
\end{center}
\end{table}
\begin{eqnarray}
M^{\pi}&=& \langle \pi (\vec{p}), B_{Q2}(v) \mid T \mid B_{Q1}(v) \rangle
\nonumber\\[2mm]
&=&{\bar \psi}_{2}^{\nu_{1} \dots \nu_{j_{2}}}(v) \psi_{1}^{\mu_{1} \dots
\mu_{j_{1}}}(v) (\sum_{l_{\pi}} f_{l_{\pi}} t^{l_{\pi}}_{\mu_{1} \dots
\mu_{j_{1}};\nu_{1} \dots \nu_{j_{2}}}) \,.\label{trans}
\end{eqnarray}
The light diquark transition tensors
$t^{l_{\pi}}_{\mu_{1} \dots \mu_{j_{1}};\nu_{1} \dots \nu_{j_{2}}}$
of rank $(j_{1}+j_{2})$, built from
$g_{\perp\mu\nu}=g_{\mu\nu}-v_{\mu}v_{\nu}$ and
$p_{\perp\mu}=p_{\mu}-v\cdot pv_{\mu}$, should have the correct parity and
project out
the appropriate partial wave amplitude with amplitude $f_l$. From now on, we
shall sometimes use $l$ to refer to the pion momentum $l_{\pi}$ if there
can be no confusion.
  In Table \ref{tabtens} we summarize the relevant covariant tensors for the
  allowed
  diquark transitions considered in this paper. We have introduced the
  traceless, symmetric and second
  rank tensor $T_{\mu\nu}$, appropriate for d-wave transitions
  from {\it P} wave to {\it S} wave. It is defined by
\be
T_{\mu_1\mu_2}(p)=p_{\perp \mu_1}p_{\perp \mu_2}-
\frac{p_{\perp}^2}{3}g_{\perp \mu_1 \mu_2}\;\;,
\ee
note that $p_{\perp}^2=-|vec{p}^2|$.
One also needs the corresponding third rank and traceless tensor
$T_{\mu\nu\rho}$ appropriate for f-wave transitions among the
{\it P} wave states which is given by
\be\label{T-f}
T_{\mu_1\mu_2\mu_3}=p_{\perp \mu_1}p_{\perp \mu_2}p_{\perp\mu_3}-
\frac{p_{\perp}^2 }{5}\left(g_{\perp\mu_1\mu_2} 
p_{\perp \mu_3}+cycl.(\mu_1\mu_2\mu_3)\right)\;,
\ee
The f-wave tensor $T_{\mu\nu\rho}$ is symmetric with respect to the
exchange of any pair of
Lorentz indices. The construction of the most general traceless tensor
$T^{\mu_1\mu_2\cdots\mu_l}$ of rank $l$, necessary to describe
$l$-wave transitions can be done in two stages.
 One begins with projecting $T^{\mu_1\mu_2\cdots\mu_l}$  on the space
of completely symmetric
tensors by
symmetrizing it. Then one subtracts appropriate tensors in order that the
total trace is
 null. The general form of such a tensor is given in \cite{htk}.

A simple $LS$- coupling exercise shows that the number of independent
covariants or partial waves contributing to
a given transition $j_1^P\rightarrow j_2^P+\pi$ is given by  
$N=j_{min}+\frac{1}{2}(1-n_1n_2)$ where $j_{min}=min\{j_1,j_2\}$.
The normality $n$ of a diquark state with quantum numbers $j^P$ is
defined by $n=P(-1)^j$. By similar reasoning one finds that the
$\epsilon$-tensor should be present when the product of the normalities
is even ($n_1n_2=+1$). 

In Table \ref{tabhqs}, we list the allowed transitions between the heavy 
baryon ground states
({\it S} wave to {\it S} wave) and those from the
excited ({\it P} wave) state down to the ground state.
The {\it S} wave to {\it S} wave transitions involve two p-wave coupling
constants while each of the single pion
transitions from the K-multiplet and from the k-multiplet down to the ground
state are determined in terms
of seven coupling constants. There are three s-wave and four d-wave
couplings for each.

The HQS single pion transitions of Table \ref{tabhqs} are
expressed in terms of transition amplitudes. They could have equally
well be written down using the language of effective Lagrangians.
In this case the heavy baryon spin wave functions would be replaced by the
appropriate heavy baryon superfields with derivative couplings to the pion
field. 

Let us also briefly comment on the relation of our approach to the 
chiral invariant coupling method used in \cite{cho,huang,yp} when the chiral
invariant Lagrangian is expanded to first order in the pion field.   
 The chiral formalism
 implies that all the one pion coupling factors are proportional to
 the factor $1/f_\pi$  associated with the pion field.
 For the s-wave transitions in the chiral approach there is an extra
 factor $v\cdot p=E_{\pi}$, which comes in because the pion field is
 coupled via the
 scalar term $v\cdot A$, where $A$ is the nonlinear axial 
 field. Thus, in the rest frame of the heavy baryons the pion field couples
 through the scalar component of the axial vector current in the $s$-wave
 transitions. This is in contrast to our effective coupling approach, to be 
 discussed later on, where the
 coupling is always through the three-vector
 piece of the axial vector current. Also the $p_{\perp}^2$
 factors required in the construction of our d-wave and f-wave
 tensors, Table \ref{tabtens},
 can be written as $E_{\pi}^2$ in the chiral
 formalism when terms linear in the light quark masses are
 neglected (see \cite{yp}). In fact, when these terms are kept, one does
 obtain the correct $p_{\perp}^2$ factor. Up to the mentioned factors
 there is a one to one correspondence between our coupling factors and
 those in the chiral approach. 


\begin{table}
\caption{\label{tabhqs} Heavy Quark Symmetry (HQS) predictions for the
allowed single pion transitions between ground states ({\it S} wave) and
from excited
({\it P} wave) down to the ground state. }
\vspace{3mm}
\renewcommand{\baselinestretch}{1.2}
\small \normalsize
\begin{center}
\begin{tabular}{|clcl||}
\hline \hline
\multicolumn{1}{c}{Heavy baryon transitions}&
\hspace*{3cm} Transition amplitudes\\
\multicolumn{1}{c}{$B_Q^{'} \rightarrow B_Q + \pi$}& $  
\hspace*{5cm}M^{\pi}$ \\
\hline \hline     
\multicolumn{1}{r}{ Ground-State }&{ } \\
\hline \hline  
\hspace*{-2cm} $\Sigma_{Q}\rightarrow \Lambda_Q$ &
$  {\bar u}_2(v)   
       \left \{
   \begin{array}{c}
       \frac{1}{\sqrt{3}} \gamma_{\perp}^{\mu} \gamma_{5} u_{1}(v)\\
       u^{\mu}_1(v)
   \end{array}      
\right \} I_1
       f_{1p}\;p_{\perp \mu}$\\    
\hline
\hspace*{-2cm} $\Sigma_{Q} \rightarrow \Sigma_Q$ & $
  \left \{ 
  \begin{array}{c}
       \frac{1}{\sqrt{3}} {\bar u}_{2}(v)\gamma_{5}\gamma_{\perp}^{\nu}\\
       {\bar u}^{\nu}_2(v)
   \end{array} \right \}    
      \left \{
   \begin{array}{c}
       \frac{1}{\sqrt{3}} \gamma_{\perp}^{\mu} \gamma_{5} u_{1}(v)\\
       u^{\mu}_1(v)
   \end{array}      
\right \} I_2
       f_{2p}\; i\varepsilon ( \mu \nu p v)      $\\ 
\hline \hline 
\multicolumn{1}{r}{ K-multiplet }&{ } \\
\hline \hline 
\hspace*{-2cm}
$\Lambda_{QK1}\rightarrow \Sigma_Q$ &
$ \left \{ 
  \begin{array}{c}
       \frac{1}{\sqrt{3}}{\bar u}_{2}(v)\gamma_{5}\gamma_{\perp}^{\nu}\\
       {\bar u}^{\nu}_2(v)
   \end{array} \right \}    
      \left \{
   \begin{array}{c}
       \frac{1}{\sqrt{3}} \gamma_{\perp}^{\mu} \gamma_{5} u_{1}(v)\\
       u^{\mu}_1(v)
   \end{array}      
\right \} I_3 \left (
        f_{1s}^{(K)} g_{\mu\nu} + f_{1d}^{(K)} T_{\mu\nu}      
\right ) $\\    
\hline
\hspace*{-2cm}
 $\Sigma_{QK0}\rightarrow \Lambda_Q$&
$I_1  f_{2s}^{(K)}{\bar u}_2(v)u_1(v) $\\
\hline
\hspace*{-2cm} $\Sigma_{QK1} \rightarrow \Sigma_Q$ & $
  \left \{ 
  \begin{array}{c}
       \frac{1}{\sqrt{3}} {\bar u}_{2}(v)\gamma_{5}\gamma_{\perp}^{\nu}\\
       {\bar u}^{\nu}_2(v)
   \end{array} \right \}    
      \left \{
   \begin{array}{c}
       \frac{1}{\sqrt{3}} \gamma_{\perp}^{\mu} \gamma_{5} u_{1}(v)\\
       u^{\mu}_1(v)
   \end{array}      
\right \} I_2 \left (
       f_{3s}^{(K)} g_{\mu\nu} + f_{3d}^{(K)} T_{\mu\nu}      
\right ) $\\  
\hline
\hspace*{-2cm} $ \Sigma_{QK2} \rightarrow\Lambda_Q$ &
        ${\bar u}_{2}(v) \left \{
   \begin{array}{c}
       \frac{1}{\sqrt{10}}\gamma_5 \gamma_{\perp}^{\{\mu_1}u_1^{\mu_2\}_0}(v)\\
       u_1^{\mu_1\mu_2}(v)
   \end{array}      
\right \} I_1 f_{4d}^{(K)} T_{\mu_1\mu_2}  $\\
\hline
\hspace*{-2cm} $\Sigma_{QK2}\rightarrow \Sigma_Q$ &
$ \left \{ 
  \begin{array}{c}
       \frac{1}{\sqrt{3}}{\bar u}_{2}(v) \gamma_{5}\gamma_{\perp}^{\nu}  \\
       {\bar u}^{\nu}_2(v)
   \end{array} \right \} \left \{
   \begin{array}{c}
       \frac{1}{\sqrt{10}}\gamma_5 \gamma_{\perp}^{\{\mu_1}u_1^{\mu_2\}_0}(v)\\
       u_1^{\mu_1\mu_2}(v)
   \end{array}      
\right \} I_2 f_{5d}^{(K)} i\varepsilon_{\mu_1 \nu \sigma \rho}v^{\sigma}T^{\rho}_{\mu_2} $\\
\hline \hline     
\multicolumn{1}{r}{ k-multiplet }&{ } \\
\hline \hline  
\hspace*{-2cm} $\Sigma_{Qk1}\rightarrow \Sigma_Q$ &
$ \left \{ 
  \begin{array}{c}
       \frac{1}{\sqrt{3}} {\bar u}_{2}(v)\gamma_{5}\gamma_{\perp}^{\nu}\\
       {\bar u}^{\nu}_2(v)
   \end{array} \right \}    
      \left \{
   \begin{array}{c}
       \frac{1}{\sqrt{3}} \gamma_{\perp}^{\mu} \gamma_{5} u_{1}(v)\\
       u^{\mu}_1(v)
   \end{array}      
\right \} I_2 \left (
       f_{1s}^{(k)} g_{\mu\nu} + f_{1d}^{(k)} T_{\mu\nu}      
\right ) $\\    
\hline
\hspace*{-2cm} $\Lambda_{Qk0}\rightarrow \Lambda_Q$ &
$ {\bar u}_{2}(v)u_1(v)  
 I_4 f_{2s}^{(k)}  $\\    
\hline
\hspace*{-2cm} $\Lambda_{Qk1} \rightarrow \Sigma_Q$ & $
  \left \{ 
  \begin{array}{c}
       \frac{1}{\sqrt{3}} {\bar u}_{2}(v)\gamma_{5}\gamma_{\perp}^{\nu}\\
       {\bar u}^{\nu}_2(v)
   \end{array} \right \}    
      \left \{
   \begin{array}{c}
       \frac{1}{\sqrt{3}} \gamma_{\perp}^{\mu} \gamma_{5} u_{1}(v)\\
       u^{\mu}_1(v)
   \end{array}      
\right \} I_3 \left (
       f_{3s}^{(k)} g_{\mu\nu} + f_{3d}^{(k)} T_{\mu\nu}      
\right ) $\\  
\hline
\hspace*{-2cm}$\Lambda_{Qk2}\rightarrow \Lambda_Q$ &
$ {\bar u}_{2}(v)  \left \{
   \begin{array}{c}
       \frac{1}{\sqrt{10}}\gamma_5 \gamma_{\perp}^{\{\mu_1}u_1^{\mu_2\}_0}(v)\\
       u_1^{\mu_1\mu_2}(v)
   \end{array}      
\right \} I_4   f_{4d}^{(k)} T_{\mu_1\mu_2}  
$\\  
\hline
\hspace*{-2cm}$\Lambda_{Qk2}\rightarrow \Sigma_Q$ &
$ \left \{ 
  \begin{array}{c}
       \frac{1}{\sqrt{3}}{\bar u}_{2}(v) \gamma_{5}\gamma_{\perp}^{\nu}  \\
       {\bar u}^{\nu}_2(v)
   \end{array} \right \} \left \{
   \begin{array}{c}
       \frac{1}{\sqrt{10}}\gamma_5 \gamma_{\perp}^{\{\mu_1}u_1^{\mu_2\}_0}(v)\\
       u_1^{\mu_1\mu_2}(v)
   \end{array}      
\right \} I_3 f_{5d}^{(k)} i\varepsilon_ { \mu_1 \nu \rho \sigma}v^{\sigma}T^{\rho}_{\mu_2}
$\\  
\hline \hline            
\end{tabular}
\end{center}
\renewcommand{\baselinestretch}{1}
\small \normalsize
\end{table}

The decay rates of the various transitions of Table \ref{tabhqs} can be
calculated using the general rate formula
\be\label{rate}
\Gamma = \frac{1}{2J_{1}+1} \quad \frac{ \mid \vec{p} \mid}{8 \pi
M_{1}^{2}}\sum_{spins} \mid M^{\pi} \mid^{2}.
\ee
Here, $ \mid \vec{ p} \mid $ is the pion momentum in the heavy
baryon's rest frame $ \mid \vec{ p} \mid^2 =
p_{\perp \mu} p_{\perp \nu }g^{\mu\nu}$ and $M^{\pi}$ are the transition
amplitudes listed in Table \ref{tabhqs}.  When calculating decay rates
in terms of the covariant expressions of Table \ref{tabhqs} it is highly
advisable to use the accelerated spin sum algorithm of \cite{bf}. For example,
the four transitions in $\Lambda_{Qk2} \rightarrow \Sigma_Q +\pi $
or $\Sigma_{QK2} \rightarrow \Sigma_Q +\pi $ are given by
\begin{eqnarray}\label{rates}
\Gamma\left({\textstyle \frac{3}{2}^{-}}
\rightarrow {\textstyle \frac{1}{2}^{+}} +\pi\right) &=&f_{5d}^2
      \frac{\mid \vec{p} \mid^{5}}{20\pi}\frac{M_{2}}{M_{1}} \nonumber \\
\Gamma\left({\textstyle \frac{3}{2}^{-}}
\rightarrow {\textstyle \frac{3}{2}^{+}} +\pi\right) &=&f_{5d}^2
      \frac{\mid \vec{p} \mid^{5}}{20\pi}\frac{M_{2}}{M_{1}} \nonumber \\
\Gamma\left({\textstyle \frac{5}{2}^{-}}
\rightarrow {\textstyle \frac{1}{2}^{+}} +\pi\right) &=&f_{5d}^2
      \frac{1 \mid \vec{p} \mid^{5}}{45\pi}\frac{M_{2}}{M_{1}} \nonumber \\
\Gamma\left({\textstyle \frac{5}{2}^{-}}
\rightarrow {\textstyle \frac{3}{2}^{+}} +\pi\right) &=&f_{5d}^2
      \frac{7\mid \vec{p} \mid^{5}}{90\pi}\frac{M_{2}}{M_{1}}
\end{eqnarray}
where the flavor factors have been omitted.
It is easy, from these equations, to show that the decay rates for 
these four one--pion transitions satisfy the following ratios 
\be\label{ratio of rates1}
\Gamma_{\frac{3}{2}^{-}\rightarrow \frac{1}{2}^{+}+\pi}:
\Gamma_{\frac{3}{2}^{-}\rightarrow \frac{3}{2}^{+}+\pi}:
\Gamma_{\frac{5}{2}^{-}\rightarrow \frac{1}{2}^{+}+\pi}:
\Gamma_{\frac{5}{2}^{-}\rightarrow \frac{3}{2}^{+}+\pi}
=9:9:4:14.
\ee
Moreover, we have
\be\label{ratio of rates2}
\Gamma_{\frac{3}{2}^{-}\rightarrow \frac{1}{2}^{+}+\pi}+
\Gamma_{\frac{3}{2}^{-}\rightarrow \frac{3}{2}^{+}+\pi}=
\Gamma_{\frac{5}{2}^{-}\rightarrow \frac{1}{2}^{+}+\pi}+
\Gamma_{\frac{5}{2}^{-}\rightarrow \frac{3}{2}^{+}+\pi}.
\ee
The result, Eq.(\ref{ratio of rates1}), and the sum rule,
 Eq.(\ref{ratio of rates2}), agree with
the conventional approach using Clebsch--Gordan coefficients \cite{IW91}
and also with the more recent and compact procedure using the 6j-symbols
\cite{kkp,Z93,fp93}.
 In fact, using the 6j-symbol approach, one can write down a closed
form expression for the decay rates of all transitions. One has
\[
\Gamma_l^i = \frac{1}{\pi} \frac{M_2}{M_1} \mid \vec{p}
\mid^{2l+1} \mid f_{il} \mid^2 |I_i|^2 \mid c(j_1,j_2,l)\mid^2 \]
\be\label{rate-gen}
(2j_1+1)(2J_2+1)\biggl( \biggl\{
    \begin{array}{rrr}
     l& j_1& j_2 \\
     s_Q & J_2 & J_1 \\
     \end{array}
    \biggr\} \biggr)^2\;.
\ee
 Here, we have included the flavor factors $|I_i|^2$ which depend on the
 specific flavor channels involved in the transition. 
Explicit forms for these SU(3) flavor factors are given in 
Eq.(\ref{I-factors}).
The curly bracket stands
for the usual
6j-symbol given in Table \ref{tab6j} of appendix B and $c(j_1,j_2,l)$ is 
the ratio of the 
reduced amplitude appearing
in the 6j-approach and the invariant coupling factor $f_{il}$ of
Table \ref{tabhqs}. This proportionality factor is a function of $j_1$,
$j_2$ and $l$ alone. The rate factors $c(j_1,j_2,l)$  
for the transitions discussed in this paper are given in Table \ref{tabtens}.

Now, using the standard orthogonality relation for the 6j-symbols
\be\label{6j-sumrule}
\sum_{J_2} (2j_1+1)(2J_2+1)
\biggl(\left\{
    \begin{array}{rrr}
     l& j_2& j_1 \\
     \frac{1}{2} & J_1 & J_2 \\
     \end{array}
    \right\}\biggr)^2 =1  \;\;,
\ee
 one can immediately derive Eq.(\ref{ratio of rates2}). This sum rule 
 shows that the total rate of pionic decays from any of the two HQS
doublet states $J_1=j_1\pm \frac{1}{2}$ into the HQS doublet states
$J_2=j_2\pm \frac{1}{2}$ is independent of $J_1$.  In a similar
manner one concludes that, the rates of transition into a heavy quark
singlet state from the two doublet states and vice versa are identical
to one another. 

Table \ref{tabhqs} also contains the appropriate $SU(3)$ factors denoted by
\begin{eqnarray}\label{I-factors} 
I_1\;\;(6\rightarrow 3^{*}+\pi)&=&T^{(3^{*})ac}T^{(6)}_{bc}{\bar M}_a^b
\nonumber\\
I_2\;\;(6\rightarrow 6+\pi)&=&T^{(6)ac}T^{(6)}_{bc}{\bar M}_a^b \nonumber  \\
I_3\;\;(3^{*}\rightarrow 6+\pi)&=&T^{(6)ac}T^{(3^{*})}_{bc}{\bar M}_a^b      \\
I_4\;\;(3^{*}\rightarrow 3^{*}+\pi)&=&T^{(3^{*})ac}T^{(3^{*})}_{bc}{\bar
M}_a^b\;\;,
\nonumber 
\end{eqnarray}
where $T^{(6)}_{ab}$ and  $T^{(3^{*})}_{ab}$ are respectively the symmetric
and antisymmetric flavor tensors, transforming as sextet and anti-triplet in
$SU(3)$ and are built from two light quark states. Explicit expressions for
these tensors are given by: \\
i)$ \;\;\;\; T^{(6)}_{ab}$
\begin{eqnarray} 
\Sigma_c^{++} \;\;\;\; &:& \;\;\;\; uu \nonumber \\ 
\Sigma_c^{+} \;\;\;\; &:& \;\;\;\; \frac{1}{\sqrt{2}}(ud+du)  \nonumber \\ 
\Sigma_c^{0} \;\;\;\; &:& \;\;\;\; dd \nonumber  \\ 
\Xi_c^{' +} \;\;\;\; &:& \;\;\;\; \frac{1}{\sqrt{2}}(us+su)   \\
\Xi_c^{' 0} \;\;\;\; &:& \;\;\;\; \frac{1}{\sqrt{2}}(ds+sd) \nonumber \\
\Omega_c^{0} \;\;\;\; &:& \;\;\;\; ss \nonumber    
\end{eqnarray}             
ii)$\;\;\;\; T^{(3^{*})}_{ab}$
\begin{eqnarray} 
\Lambda_c^{+} \;\;\;\; &:& \;\;\;\; \frac{1}{\sqrt{2}}(ud-du)  \nonumber  \\ 
\Xi_c^{+} \;\;\;\; &:& \;\;\;\; \frac{1}{\sqrt{2}}(us-su)  \\
\Xi_c^{0} \;\;\;\; &:& \;\;\;\; \frac{1}{\sqrt{2}}(ds-sd) \nonumber   
\end{eqnarray}             
Note that we have labeled the members of the sextet and anti-triplet
representations
\footnote{ We normalize the $T^{(3^{*})}_{ab}$ states to unity while in
Ref. \cite{yp} they are normalized to $2$. Therefore, the relations between
the $SU(3)$ factors $I^{'}_i$ defined
 in  \cite{yp} to ours are $I^{'}_1=\sqrt{2}I_1$,
$I^{'}_3=\sqrt{2}I_3$ and $I^{'}_4=2I_4$. } 
according to their charm
baryon content. In the bottom sector one would have to make the changes
$c\rightarrow b$ and lower the respective charge by $1$. ${\bar M}_a^b$ is the
transpose of the $3\times 3$  pion flavor wave function with the components
$\pi^{+}:u{\bar d}$ , $\pi^{0}:\frac{1}{\sqrt{2}}(u{\bar u}-d{\bar d})$ and
 $\pi^{-}:d{\bar u}$.
\section{Constituent Quark Model Relations}
To describe the light diquark transitions, we shall use the 
constituent quark model with its underlying $SU(6)\otimes O(3)$ symmetry       
to go beyond the Heavy Quark Symmetry predictions.
For the {\it S} wave to {\it S} wave transitions, the light diquark tensors
of the
matrix elements in Eq. (\ref{trans}), including flavor factors, can be written as
\be\label{two}
t^{l_{\pi}}_{\mu_{1} \dots
\mu_{j_{1}};\nu_{1} \dots \nu_{j_{2}}}=\left(\bar{\hat \phi}_{\nu_1 \dots \nu_{j_2}}\right)^{AB}
\left({\cal O}^{l_{\pi}}\right)^{A^{'}B^{'}}_{AB}
\left({\hat \phi}_{\mu_1 \dots \mu_{j_1}}\right)_{A^{'}B^{'}} \; .  
\ee                                                                          
Here, the light diquark spin wave functions $\hat \phi$'s involve either
${\chi}^0_{\alpha\beta}T_{ab}^{3^{*}}$ 
for spin zero or ${\chi}^{1\mu}_{\alpha\beta}T_{ab}^{6}$ for spin one and
the operator ${\cal O}$ is given by an overlap integral which we do not
attempt to calculate. We have retained the generic representation of the
light diquark spin wave functions in terms of $j_1$ and $j_2$ Lorentz indices.
For {\it S} wave to {\it S} wave pion transitions, the partial wave
involved is
$l_{\pi}=1$ and hence it is easy to see that $\cal O$ must be a
pseudoscalar operator involving one power of the pion momentum $p$. In the
constituent quark model the pion is emitted by just one of the light quarks,
therefore, the transition operator ${\cal O}$  
must be a one-body operator. In $1/N_C$ \cite{witten}, two-body emission 
operators for pion couplings to {\it S} wave heavy baryons are non leading
and 
can be neglected in the constituent quark model approach \cite{cgkm,py-two}.
The $1/N_C$-approach thus provides a justification of   
neglecting two-body and higher emission operators in the constituent 
quark model.  

Because of the overall symmetry of the light diquark spin-flavor wave 
function
 $\left({\hat \phi}_{\mu_1 \dots \mu_{j_1}}\right)_{A^{'}B^{'}}$ for
{\it S} waves,
one then has uniquely
\be\label{otwo}
\left({\cal O}(p)\right)^{AB}_{A^{'}B^{'}} =\frac{1}{2}
\left((\gamma_\sigma\gamma_{5})^{A}_{A^{'}}
 \otimes (\eins)^{B}_{B^{'}}
          + \,(\eins )^{A}_{A^{'}}\otimes (\gamma_\sigma\gamma_5)
^{B}_{B^{'}}\right)
 f_{p}\,p_{ \perp}^{\sigma}\,
\ee
with $(\gamma_\sigma\gamma_{5})^{A}_{A^{'}}$ defined as
\be
(\gamma_\sigma\gamma_{5})^{A}_{A^{'}}=
(\gamma_\sigma\gamma_{5})^{\alpha}_{\alpha^{'}}{\bar M}^{a}_{a^{'}}\,.
\ee
                                           
 When considering excited states, it is convenient to explicitly pull out the
relative momentum factor in the light diquark spin wave functions according
to
\be
{\hat \phi}^{\mu_1 \dots \mu_{j}}(v,p)={\hat \phi}^{\mu_1 \dots \mu_{j};\lambda}(v)
p_{\perp\lambda}
\ee
where $p_{\perp\lambda}$ is either $k_{\perp\lambda}$ or $K_{\perp\lambda}$ and
$ {\hat \phi}^{\mu_1 \dots \mu_{j};\lambda}(v)$ is only a function of $v$.
The generalization to higher orbital excitations is straightforward. The
relevant matrix elements for {\it P} wave to {\it S} wave transitions are
given
by
\be\label{ptos}
 t^{l_{\pi}}_{\mu_{1} \dots
\mu_{j_{1}};\nu_{1} \dots \nu_{j_{2}}}=
\left(\bar{\hat \phi}_{\nu_1 \dots \nu_{j_2}}\right)^{AB}
\left({\cal O}_{\lambda}^{(l_{\pi})}\right)
^{A^{'}B^{'}}_
{AB}\left({\hat \phi}_{\mu_1 \dots \mu_{j_1};\lambda}
\right)_{A^{'}B^{'}}\,.  
\ee                                                                          
Here, the operators ${\cal O}_{\lambda}^{(l_{\pi})}$ are responsible for the
s-wave ($l_{\pi}=0$) and d-wave ($l_{\pi}=2$) transitions. Neglecting
two-body emission contributions, the transition operators 
from the K-multiplet down to the ground state are given by
\be
\left({\cal O}^{K}_{\lambda}(p)\right)^{AB}_{A^{'}B^{'}} =
\frac{1}{2}\left((\gamma^\sigma\gamma_5)^{A}_{A^{'}}
 \otimes (\eins)^{B}_{B^{'}}
          + \,(\eins )^{A}_{A^{'}}\otimes (\gamma^\sigma\gamma_5)
^{B}_{B^{'}}\right)
\left( f_s^{(K)}\;g_{\sigma\lambda}+ f_d^{(K)}\;T_{\sigma\lambda}\right)\; .                  
\label{Ktrans}
\ee
For transitions from the k-multiplet the operators instead have
the form
\be
\left({\cal O}^{k}_{\lambda}(p)\right)^{AB}_{A^{'}B^{'}} =
\frac{1}{2}\left((\gamma^\sigma\gamma_5)^{A}_{A^{'}}
 \otimes (\eins)^{B}_{B^{'}}
          - \,(\eins )^{A}_{A^{'}}\otimes (\gamma^\sigma\gamma_5)
^{B}_{B^{'}}\right)
\left( f_s^{(k)}\;g_{\sigma\lambda}+ f_d^{(k)}\;T_{\sigma\lambda}\right)\,.                  
\label{ktrans}
\ee
The relative minus sign in the effective operator Eq. (\ref{ktrans})
comes about because the
quark-quark operator is inserted between the orbitally antisymmetric
$k$-states and the orbitally symmetric $S$-wave states. In the 
$1/N_C$-approach, the contributions of one- and two-body emission operators 
are of the same order when $P$-wave baryons are involved in pion
transitions. However, their contributions are proportional to one another 
so that one needs to keep only the one-body emission operators in the 
constituent quark model approach \cite{py-two}.
 
One can also proceed to construct the operators
${\cal O}_{\lambda_1 \lambda_2}^{(l_{\pi})}$ for the 
allowed {\it P} wave to {\it P} wave transitions
associated with the p-wave ($l_{\pi}=1$) and f-wave ($l_{\pi}=3$). We write
\[
\left({\cal O}_{\lambda_1\lambda_2}(p)\right)^{AB}_{A^{'}B^{'}} =
\frac{1}{2}\left((\gamma^\sigma\gamma_5)^{A}_{A^{'}}
 \otimes (\eins)^{B}_{B^{'}}
          \pm \,(\eins )^{A}_{A^{'}}\otimes (\gamma^\sigma\gamma_5)
^{B}_{B^{'}} \right)  \]
\be\label{omunu}
(\sum_{L=0}^{2} g^{(L)}_{p}P^{(L)}_{\sigma\lambda_1\lambda_2}+ 
g_{f}T_{\sigma\lambda_1\lambda_2} ) \; ,      
\ee
with 
\begin{eqnarray}
P^{(0)}_{\sigma\lambda_1\lambda_2}&=&p_{\perp \sigma}g_{\lambda_1\lambda_2} 
\nonumber \\
P^{(1)}_{\sigma\lambda_1\lambda_2}&=&\frac{1}{2}(p_{\perp\lambda_1}
g_{ \sigma\lambda_2}-p_{\perp\lambda_2}g_{ \sigma\lambda_1} ) \nonumber \\
P^{(2)}_{\sigma\lambda_1\lambda_2}&=&\frac{1}{2}(p_{\perp\lambda_1}
g_{ \sigma\lambda_2}-p_{\perp\lambda_2}g_{ \sigma\lambda_1} )-
\frac{1}{3}p_{\perp\sigma}g_{ \lambda_1\lambda_2} \label{T-tensors} \; ,
\end{eqnarray}
and the third rank tensor $T_{\sigma\lambda_1\lambda_2}$ is defined 
in Eq. (\ref{T-f}). The $p$-wave transition tensors
$P^{(L)}_{\sigma\lambda_1\lambda_2}$, with $L=0,1, {\rm and\;} 2$ correspond 
to the three possible angular momenta inducing the transitions between the
two orbital {\it P} wave states. The p-wave and f-wave coupling constants
are
represented by $g^{(L)}_{p}$ and $g_f$, respectively. Note that, from
angular momentum conservation, there is only one coupling possibility
for the f-wave transition. In Eq.
(\ref{omunu}), the $(+)$ sign has to be used for the $K\rightarrow K$ and
$k\rightarrow k$ pionic transitions, whereas the $(-)$ sign is appropriate 
for the $K\rightarrow k$ and the $k\rightarrow K$ transitions. 
The constituent quark model analysis of {\it P} wave to 
{\it P} wave transitions will be presented in Appendix A.
 The generalization to transitions involving higher orbital excitations
is straightforward.

The matrix elements Eq. (\ref{two}) and Eq. (\ref{ptos}) of the
operators Eq. (\ref{otwo}),  Eq. (\ref{Ktrans}) and Eq. (\ref{ktrans}), can
be readily evaluated using the light diquark
spin wave functions in Table \ref{tab1}. The two ground state to ground state
coupling strengths can be seen to be related to the single coupling $f_{p}$ as

\be\label{fp}
f_{1p}=f_{2p}=f_p \, .
\ee
For {\it P} wave (K-multiplet) to {\it S} wave transitions the evaluation
of the
matrix
elements leads to the following relations 
\begin{eqnarray}\label{fsK}
f_{1s}^{(K)}=f_s^{(K)}\;\; ; \;\; f_{2s}^{(K)}=\sqrt{3}f_s^{(K)}
\;\; ; \;\;f_{3s}^{(K)}=-\sqrt{2}f_s^{(K)} \\
f_{1d}^{(K)}=f_d^{(K)}\;\; ; \;\; f_{3d}^{(K)}=\frac{1}{\sqrt{2}}f_d^{(K)}
\;\; ; \;\;  f_{4d}^{(K)}=f_d^{(K)}\;\; ; \;\;  f_{5d}^{(K)}=-f_d^{(K)}
\label{fdK}
\end{eqnarray}      
The number
of independent coupling constants has been reduced from seven to the two
constituent quark model s-wave and d-wave coupling
factors $f_{s}^{(K)}$ and $f_{d}^{(K)}$.

In a similar way, one reduces the seven decay couplings for
pion transitions from the excited k-multiplet to the two corresponding
constituent quark model coupling factors 
$f_{s}^{k}$ and $f_{d}^{k}$. With the appropriate replacement
\be\label{fk}
f_{is,id}^{(K)}\rightarrow f_{is,id}^{(k)} \;{\rm and}\; 
f_{s,d}^{(K)}\rightarrow f_{s,d}^{(k)} \; ,
\ee 
these relations are identical to those in Eqs. (\ref{fsK}) and (\ref{fdK}). 
The constituent quark model predictions for {\it P} wave to {\it S} wave
are in
agreement, 
after taking care of the different normalization of the $\Lambda$-type
flavor factors, with corresponding results in the Chiral formalism 
\cite{yp}. 
We mention that the constituent quark model
calculation in \cite{yp} has been done using static quark model
wave functions and explicit Clebsch-Gordan coefficients as compared
to our covariant approach. The two methods are of course equivalent to
each other even if they use different calculationl techniques.  

In appendix A, we give our results on the {\it P} wave to {\it P} wave
single pion
transitions. It is obvious that the {\it D} wave to {\it S} wave or {\it D} 
wave
to
{\it P} 
wave transitions, not treated here, can be analyzed
following the same procedure. 
\section{Phenomenological Predictions}
Pion transitions of charm baryons are interesting from the
experimental point of view where data are already published by some
laboratories \cite{exp,cleo}.
Therefore, it is important to predict some of these
transition rates and to compare them with other theoretical models. 
The constituent quark model coupling $f_p$ determines the single pion 
transitions among the {\it S} wave heavy baryon states. On the other 
hand, $f^{(K)}_s$ and $f^{(K)}_d$ are sufficient to predict transitions 
from the {\it P} wave K-multiplet to the ground state.

Using PCAC the p-wave coupling constant $f_p$ in Eq. (\ref{fp}) can be 
related to the quark's  axial vector current coupling 
strength $g_A$ \cite{ycclly,cheng}, one obtains
\be\label{fp-PCAC}
f_{p}=g_A/f_{\pi}\; ,
\ee
with $g_A$ of the order of unity. However, we mention  that there is some 
additional theoretical support for values of $g_A < 1$  from an analysis 
of the NJL model \cite{bijnens} and the Skyrme model \cite{glm}. 
This result is also obtained if one demands that the 
experimentally measured $G_A/G_V$ value comes out right in the constituent 
quark model. An indication about the $f_{p}$ strength can be obtained from 
Eq. (\ref{fp-PCAC}) by taking $g_A \approx 0.75 $ \cite{ycclly,cheng} 
and $f_{\pi}=0.093 {\rm \; GeV}$. One gets
\be
f_{p}=8.06  {\rm \; GeV^{-1}}.
\ee 
The single-pion decay
rates can be calculated using the rate formula, Eq.(\ref{rate}), 
Table \ref{tabhqs} and the relations (\ref{fp}) and (\ref{fsK}-\ref{fk}).
Considering {\it P} wave to {\it S} wave transitions,   
the decay rates for the one pion transition of the $\Lambda_{QK1}$
doublet to $\Sigma_c$ are given by 
\begin{eqnarray}\label{srates}
\Gamma\left({\textstyle \frac{1}{2}^{-}}
\rightarrow {\textstyle \frac{1}{2}^{+}} +\pi\right) &=&{f_{s}^{{(K)}^2}}I_3^2
      \frac{\mid \vec{p} \mid}{2\pi} \frac{M_{2}}{M_{1}}\label{rate2}\\
\Gamma\left({\textstyle \frac{3}{2}^{-}}
\rightarrow {\textstyle \frac{1}{2}^{+}} +\pi\right) &=&{f_{d}^{{(K)}^2}}I_3^2
      \frac{\mid \vec{p} \mid^{5}}{18\pi}\frac{M_{2}}{M_{1}} \label{rate3} \; .
\end{eqnarray}       
These equations are derived using Eq (\ref{rate-gen}) and Table
\ref{tabtens}.
Replacing $f_s^{(K)}$ by $f_s^{(k)}$, $f_d^{(K)}$ by $f_d^{(k)}$ 
and $I_3$ by $I_2$, one can
immediately calculate the decay rates for the one pion transition 
$\Sigma_{Qk1}\rightarrow\Sigma_c$ for which there is unfortunately no
data available at present.
The other interesting pion transitions are those for
$\Sigma_{QK2} \rightarrow \Sigma+\pi$ and 
$\Lambda_{Qk2} \rightarrow \Sigma_c+\pi$ which proceed via d-wave 
transitions.
In terms of $J^{P}$ quantum numbers one has the 
transitions $\{ \frac{3}{2}^{-},
\frac{5}{2}^{-}\} \rightarrow \{ \frac{1}{2}^{+}, \frac{3}{2}^{+} \} + \pi$
and their decay rates are given by Eq. (\ref{rates}).

 Using $SU(6)\times O(3)$ symmetry the coupling
constants $f_s^{(K)}$ and $f_d^{(K)}$ are
sufficient to predict transitions from the K-multiplet to the ground state.
 To estimate the $f_s^{(K)}$ coupling, 
  we use the recent experimental values for the decay width of
  $\Lambda_{cK1}(\frac{1}{2}^-)$ ($\Gamma_{\Lambda_{cK1}(2593)}=
  3.6^{+3.7}_{-3.0} $ MeV)
  reported by CLEO \cite{cleo}. Assuming that this width is saturated by the
  strong decay and using Eq.(\ref{rate2}) one gets
 \be\label{fs1}
  f_s^{(K)}=1.05^{+0.54}_{-0.44}
  \ee
   The uncertainty in $f_s^{(K)}$ is mainly due to the experimental
  error in $\Gamma_{\Lambda_{cK1}(\frac{1}{2}^-)}$. To predict the
  value of the coupling constant in the Chiral formalism $h_2$
 defined in \cite{cf,cho,huang,yp} we use 
 $h_2=f_s^{(K)}\frac{f_{\pi}}{E_{\pi}}$, with $f_{\pi}=0.093 {\rm \; GeV}$, to
  obtain
 \be
 h_2=0.69^{+0.35}_{-0.29 } \; .
 \ee
 Strong decays involving $\Xi^{* +}_{cK1}$ can also be used to determine
the 
 coupling $f_s^{(K)}$. Assuming that $70\%$ of the $\Xi^{* +}_{cK1}(2815)$ 
 width is saturated by $\Xi^{* +}_{cK1}(2815)\rightarrow \Xi^{*
0}\pi^{+}$,
  one finds  
 \be\label{fs2}
 f_s^{(K)}=0.48 \; ,
\ee 
  here we have taken ($\Gamma_{\Xi^{*}_{cK1}(\frac{3}{2}^+)}<3 {\rm \;
MeV}$) 
  which is the upper limit set by CLEO \cite{cleo}. This result
 suggests that the strength of the $\Lambda_{cK1}$ single 
 pion coupling to $\Sigma_c$ is about twice the $\Xi^{*}_{cK1}$ to 
 $\Xi^{*}_c$ coupling. We would like to mention that most our predictions
are in agreement with those reported in ref. \cite{3-quark} using a
three-quak model.  
  
 There is no precise experimental data available which can be used to
estimate the $f_d^{(K)}$ numerical value. 
However, the upper bound set by CLEO for the 
 transition $\Lambda^{*}_{cK1}(2625)\rightarrow\Sigma^{0}_c\pi^{+}<0.13 
 {\rm \; MeV}$ and Eq. (\ref{rate3}) can be used to predict the coupling 
 $f^{(K)}_d$. One obtains
 \be\label{fd1}
 f^{(K)}_d < 27.26\pm0.06 {\rm\;\; GeV^{-2}}.
 \ee
 One may also uses the decay
$\Lambda(1520)\rightarrow\Sigma\pi$, reported in \cite{PDG}, and 
Eq. (\ref{rate3}) to get some idea about the strength of this coupling 
which is very sensitive 
to the numerical value of the emitted pion momentum. 
Moreover, one should bear in mind that, since the strange quark
is not heavy enough, the $1/m_s$ corrections can be important in
this case. Therefore, the predicted rates and couplings should be taken only as
rough guesses. 
Using the published decay rate for this transition, the coupling 
$f^{(K)}_d$ is estimated to be
\be\label{fd2}
f^{(K)}_d=18.62\pm0.05 {\rm\;\; GeV^{-2}}.
\ee
Using this result, one predicts the one-pion decay rates for  
$\Lambda_{cK1}(2625)\rightarrow\Sigma_c$ to be
\be\label{decay-fd2}
 \Gamma_{\Lambda_{cK1}(2625)\rightarrow\Sigma^{0}_c\pi^{+}}=0.05 {\rm\;\; MeV}.
\ee
One concludes that 
the spin-$(\frac{3}{2}^-)$ member of the $\Lambda_{cK1}$ doublet is
constrained to decay to the $\Sigma_c(\frac{1}{2}^+)$ via d-wave which is
suppressed. Its preferred $s$-wave decay into $\Sigma_c(\frac{3}{2}^+)$
cannot occur because this channel is not accessible kinematically.
In fact, the SCAT group reported the first evidences for the
$\Sigma_c(\frac{3}{2}^+)$ at a mass of $2530 {\rm \; MeV}$ \cite{SCAT}. This was
confirmed later on by the CLEO Collaboration who quote a mass of
$\approx 2518 {\rm \; MeV}$ for the $\Sigma_c(\frac{3}{2}^+)$ \cite{cleo}. 
The suppression of the
$\Lambda_{cK1}(\frac{3}{2}^-)\rightarrow\Sigma_c(\frac{1}{2})$ single pion
decay mode is in
agreement with theoretical predictions reported in Ref \cite{huang}
obtained within the framework of chiral perturbation theory. 
 These
 predictions have to wait until more experimental data will be 
 available, however, it is  
still within the range of the most recent measurement published by the CLEO
collaboration \cite{cleo} 
 ($\Gamma_{\Lambda_{cK1}}(2625)<1.9 {\rm \; MeV}$).
\section{Summary and Conclusion}
 We have written down the most general one pion coupling
 structure in the heavy quark symmetry limit guided by Lorentz and
 flavor invariance. It is not difficult to see that the same coupling 
 structure emerges when considering the leading order contribution of the 
 corresponding chirally invariant Lagrangians written in
 \cite{cho,huang,yp}. Using a constituent quark model approach we exploited
 the $SU(2N_f)\times O(3)$ symmetry for the light diquark system to
 significantly reduce the number of independent 
 coupling factors. The constituent quark model predictions were worked out
 in a covariant fashion using covariant
 spin wave functions for the light diquark system. 
 Our constituent quark model predictions agree 
 with the corresponding chiral formalism \cite{yp} in which rest
 frame quark model wave functions and explicit Clebsch-Gordan 
 coefficients were used. This should not be surprising since both
 calculations are based on the same quark model picture. 
 They, in fact, must be equivalent to each other even though it is not 
 simple to see that at every step of the calculation.
 
 There is, however, a slight difference between our predictions for the 
 s-wave decay rates and those obtained by \cite{cho} and \cite{cf} who used 
 heavy quark and chiral 
 symmetries. This is due to the difference in the interaction of
the pion field and the heavy quark in the chiral formalism for S
wave transitions.
The coupling in the chiral 
 formalism is of scalar type while we use a vector coupling. This leads to the 
 appearance of an extra $(\frac{E_{\pi}}{f_{\pi}})^2$ factor in the
s-wave decay rate formula. 

 To conclude, we would like to mention that, the predictive power of the
 constituent quark model for pion transitions is limited to the heavy baryon
 sector. When applied to one pion transitions between heavy mesons
 the constituent quark model 
 provides no predictions that go beyond those of Heavy Quark Symmetry (HQS).
 In this sense heavy baryons represent an ideal setting for probing the
 dynamics of a light diquark system as we have tried to emphasize in our
 analysis.
  
\newpage
\setcounter{equation}{0}
\def\theequation{A.\arabic{equation}}
\section*{Appendix A \\ {\it P} wave to {\it P} wave one pion transitions} 
 In this appendix we shall analyze single pion transitions among the {\it
P} wave states. There
 is a proliferation of possible coupling factors for both the diagonal transitions
 ($K\rightarrow K$) and ($k\rightarrow k$) as well as for the non diagonal ones
 ($k\rightarrow K$) or
 ($K\rightarrow k$).
 In fact, one counts 8 p-wave plus 3 f-wave couplings  
 for each of the diagonal cases. For the non diagonal case one needs
 13 p-wave and 5 f-wave couplings where one should keep in mind that the
 ($k\rightarrow K$) and ($K\rightarrow k$) couplings are related to
 one another.

 In Table \ref{tabptop} we list the allowed one pion transitions among
 the {\it P} wave states  together with their
 associated coupling factors. The HQS transition
 amplitudes for the allowed
 {\it P} wave to {\it P} wave decays can be easily written down in a manner
similar
 to those quoted in Table \ref{tabhqs}. In order to save on
 space, we shall not
\begin{table}[h]
\caption{\label{tabptop}{\it P} wave to {\it P} wave allowed one pion
transitions and coupling
factors.} 
\vspace{5mm}
\renewcommand{\baselinestretch}{1.2}
\small \normalsize
\begin{center}
\begin{tabular}{|cc|cc|cc|}
\hline \hline
  $K\rightarrow K$  & coupling  & $k\rightarrow k$  &
  coupling & $k\rightarrow K$  & coupling   \\
   transitions &  factors  &  transitions &
  factors & transitions &  factors  \\
 \hline \hline
 $\Lambda_{QK1}\rightarrow \Lambda_{QK1} $ & $ g_{1p}^{(K)} $  & $\Sigma_{Qk1}\rightarrow
 \Sigma_{Qk1}$ & $ g_{1p}^{(k)}$ & $\Sigma_{Qk1}\rightarrow\Lambda_{QK1}$ & $h_{1p}$ \\
 \hspace{1cm}$\rightarrow \Sigma_{QK0} $ & $ g_{2p}^{(K)} $  &\hspace{1cm}$\rightarrow
 \Lambda_{Qk0}$ & $ g_{2p}^{(k)}$ & $\hspace{1cm}\rightarrow\Sigma_{QK0}$ & $h_{2p}$ \\
 \hspace{1cm}$\rightarrow \Sigma_{QK1} $ & $ g_{3p}^{(K)} $  &\hspace{1cm}$\rightarrow
 \Lambda_{Qk1}$ & $ g_{3p}^{(k)}$ & $\hspace{1cm}\rightarrow\Sigma_{QK1}$ & $h_{3p}$ \\
 \hspace{1cm}$\rightarrow \Sigma_{QK2} $ & $ g_{4p,4f}^{(K)} $  &\hspace{1cm}$\rightarrow
 \Lambda_{Qk2}$ & $ g_{4p,4f}^{(k)}$&$\hspace{1cm}\rightarrow\Sigma_{QK2}$ &$h_{4p,4f}$ \\
\hline
 $\Sigma_{QK0}\rightarrow \Sigma_{QK1} $ & $ g_{5p}^{(K)} $  & $\Lambda_{Qk0}\rightarrow
 \Lambda_{Qk1}$ & $ g_{5p}^{(k)}$ & $\Lambda_{Qk0}\rightarrow\Lambda_{QK1}$ & $h_{5p}$ \\
 &  &   &   & $\hspace{1cm}\rightarrow\Sigma_{QK1}$ & $h_{6p}$ \\
 \hline
 $\Sigma_{QK1}\rightarrow \Sigma_{QK1} $ & $ g_{6p}^{(K)} $ 
 & $\Lambda_{Qk1}\rightarrow
 \Lambda_{Qk1}$ & $ g_{6p}^{(k)}$ & $\Lambda_{Qk1}\rightarrow\Lambda_{QK1}$
 & $h_{7p}$ \\
 \hspace{1cm}$\rightarrow \Sigma_{QK2} $ & $ g_{7p,7f}^{(K)} $ 
 &\hspace{1cm}$\rightarrow
 \Lambda_{Qk2}$ & $ g_{7p,7f}^{(k)}$ & $\hspace{1cm}\rightarrow\Sigma_{QK0}$
 & $h_{8p}$ \\
  &   &   &  & $\hspace{1cm}\rightarrow\Sigma_{QK1}$ & $h_{9p}$ \\
  &   &   &  & $\hspace{1cm}\rightarrow\Sigma_{QK2}$ & $h_{10p,10f}$ \\
\hline
 $\Sigma_{QK2}\rightarrow \Sigma_{QK2} $ & $ g_{8p,8f}^{(K)} $ & $\Lambda_{Qk2}\rightarrow
 \Lambda_{Qk2}$&$ g_{8p,8f}^{(k)}$&$\Lambda_{Qk2}\rightarrow\Lambda_{QK1}$
 &$h_{11p,11f}$ \\
 &  &   &   & $\hspace{1cm}\rightarrow\Sigma_{QK1}$ & $h_{12p,12f}$ \\
 &  &   &   & $\hspace{1cm}\rightarrow\Sigma_{QK2}$ & $h_{13p,13f}$ \\
\hline \hline
\end{tabular}
\renewcommand{\baselinestretch}{1}
\small \normalsize
\end{center}
\end{table}
 write down the amplitudes explicitly.
 They can be constructed using Table \ref{tab1} for the heavy-side
 spin wave functions, Table \ref{tabtens}
 for the Lorentz structure and Eq.(\ref{I-factors}) for the flavor factors.
  Table \ref{tabp-p} provides three examples for the diagonal and 
  non diagonal single pion transition amplitudes among {\it P} wave states.  
 In the examples listed in Table \ref{tabp-p} we have also given the
appropriate
flavour factors according to the $SU(3)$ coupling factors of Eq.
(\ref{I-factors}). Explicitly one has the coupling factors $I_4$ for
$\Lambda
\rightarrow \Lambda$, $I_3$ for $\Lambda\rightarrow \Sigma$, $I_1$ for
$\Sigma\rightarrow \Lambda$ and $I_2$ for $\Sigma\rightarrow \Sigma$
transitions.  

 To proceed, using the constituent quark model one reduces 
 the full set of coupling factors into just three p-wave and one f-wave
coupling
 in each of the above mentioned cases. The quark model couplings
 \footnote{In order to fix the normalization of
 our coupling constants, one has to strictly adhere  
  to the use of the building blocks as prescribed above and the corresponding
  matrix elements presented in Table \ref{tabp-p} where examples for
{\it P} wave to {\it P} wave transitions are provided.}
  will be labeled as
 $g_p^{(L)}$, with $L=0,1, {\rm and \; } 2$, and $g_f$ for transitions 
 among the K-multiplet, $g_p^{\prime(L)}$ and $g_f^{\prime}$ for transitions 
 among the k-multiplet and 
 $ h_p^{(L)}$ and $ h_f$ for the non diagonal transitions. 
 They multiply the p-wave tensor $P^{(L)}_{\sigma\lambda_1\lambda_2}$ and the
 f-wave tensor $T_{\sigma\lambda_1\lambda_2}$, respectively, in the relevant 
 effective
 single pion transition operator Eq.(\ref{omunu}). 

 \begin{table}
 \caption{\label{tabp-p} Heavy Quark Symmetry (HQS) matrix elements for 
 some of {\it P} wave to {\it P} wave single pion transitions. } 
 \vspace{3mm}
 \renewcommand{\baselinestretch}{1.2}
 \small \normalsize
 \begin{center}
 \begin{tabular}{||c|c||}
 \hline \hline
  $B_Q^{'} \rightarrow B_Q + \pi$  &
  $M^{\pi}$ \\
 \hline \hline
  $\Lambda_{QK1}\rightarrow \Lambda_{QK1}$ &
 $ \left \{
  \begin{array}{c}
     \frac{1}{\sqrt{3}} {\bar u}_{2}(v)\gamma_{5}\gamma_{\perp}^{\nu}\\
	{\bar u}^{\nu}_2(v)
	\end{array} \right \}
\left \{
 \begin{array}{c}
 \frac{1}{\sqrt{3}} \gamma_{\perp}^{\mu} \gamma_{5} u_{1}(v)\\
 u^{\mu}_1(v)
   \end{array}
 \right \} I_4
  g_{1p}^{(K)}\;i\varepsilon ( \mu \nu p v) $\\
 \hline
  $\Sigma_{Qk1} \rightarrow \Sigma_{Qk1}$ & $
 \left \{
  \begin{array}{c}
   \frac{1}{\sqrt{3}} {\bar u}_{2}(v)\gamma_{5}\gamma_{\perp}^{\nu}\\
   {\bar u}^{\nu}_2(v)
\end{array} \right \}
   \left \{
 \begin{array}{c}
\frac{1}{\sqrt{3}} \gamma_{\perp}^{\mu} \gamma_{5} u_{1}(v)\\
  u^{\mu}_1(v)
 \end{array}
\right \} I_2
g_{1p}^{(k)}\; i\varepsilon ( \mu \nu p v)      $\\
\hline
  $\Sigma_{Qk1} \rightarrow \Lambda_{QK1}$ & $
 \left \{
 \begin{array}{c}
 \frac{1}{\sqrt{3}} {\bar u}_{2}(v)\gamma_{5}\gamma_{\perp}^{\nu}\\
 {\bar u}^{\nu}_2(v)
 \end{array} \right \}
  \left \{
  \begin{array}{c}
  \frac{1}{\sqrt{3}} \gamma_{\perp}^{\mu} \gamma_{5} u_{1}(v)\\
u^{\mu}_1(v)
 \end{array}
 \right \} I_1
 h_{1p}\; i\varepsilon ( \mu \nu p v)      $\\
\hline \hline
\end{tabular}
\end{center}
\renewcommand{\baselinestretch}{1}
\small \normalsize
\end{table}
 The matrix elements for the {\it P} wave to {\it P} wave single pion
transitions are given by
\[
\left(\bar{\hat \phi}_{\nu_1 \dots \nu_{j_2};\lambda_2}\right)^{AB}
\left({\cal O}_{\lambda_1 \lambda_2}^{(l_{\pi})}\right)
^{A^{'}B^{'}}_
{AB}\left({\hat \phi}_{\mu_1 \dots \mu_{j_1};\lambda_1}
\right)_{A^{'}B^{'}}\,. 
\]
Finally, using the light diquark spin wave functions in Table \ref{tab1}
and Eq.(\ref{omunu})
the heavy quark symmetry coupling factors can be related to those of the
constituent quark model. In the chiral formalism, it was shown that
the $L=1 {\;\rm and \; }2$ contributions are nominally down by two powers of
$|\vec{p} |/m$ \cite{yp}. Therefore, we shall only list the $L=0$   
case in the constituent quark model coupling factors. If needed the 
$(L=1 {\;\rm  and} \; 2)$ relations can be read off from the 
corresponding entries in Table \ref{nondiag} for the non diagonal transitions. 
We simplify the notations by dropping out the $(L=0)$ superfix and
denoting the constituent 
couplings as $g_p$ and $g_f$, 
one has the following relations\\
\hspace{0.8cm}{\bf Diagonal Transitions:}
 \[
 g_{1p}^{(K)}=0\;,\;g_{2p}^{(K)}=\frac{1}{\sqrt{3}}g_p\;,\;
 g_{3p}^{(K)}=-\frac{1}{\sqrt{2}}g_p\;,\;
 g_{4p}^{(K)}=g_p \]
 \be
 g_{5p}^{(K)}=\sqrt{\frac{2}{3}}g_p \;,
 \; g_{6p}^{(K)}=-\frac{1}{2}g_p\;,\;
 g_{7p}^{(K)}=\frac{1}{\sqrt{2}}g_p\;,\;g_{8p}^{(K)}=g_p                          
 \ee
 and
 \be
 g_{4f}^{(K)}=g_f\;,\;g_{7f}^{(K)}=-\frac{1}{\sqrt{2}}g_f\;,
 \;g_{8f}^{(K)}=g_f. 
\ee
 Similar relations hold for the ($k\rightarrow k$) transitions with the
 replacement
\be  
 g_{il}^{(K)}\rightarrow g_{il}^{(k)}  \; , \; g_p\rightarrow g^{\prime}_p  
 \; {\rm and}\;  g_f\rightarrow g^{\prime}_f  \; .
\ee
This means that diagonal transitions among {\it P} wave heavy baryon states
are described by only four constituent couplings 
$g_p$, $g^{\prime}_p$, $g_f$ and $g^{\prime}_f$. 
We mention that constituent quark model p-wave transitions have also been 
worked out in \cite{yp}, using static quark model wave functions and 
explicit Clebsch-Gordan Coefficients, which are in agreement
with our predictions. However, the results on the $f$-wave 
transitions are new. For the diagonal p-wave coupling factors, one has the
additional PCAC relations
\begin{eqnarray}
 g_p= \;\; g_p^{\prime}=\frac{g_A}{f_{\pi}}  \; .
\end{eqnarray}
\hspace{0.5cm}{\bf Non Diagonal Transitions:}\\
Finally, we work out the non diagonal case where we retain all the 
three ($L=0,1 {\; \rm and \;} 2$) p-wave transitions, which are now all nominally
of the order $(| \vec{p} |/m)^2$ since the leading contribution is zero
due to the orthogonality of the orbital wave functions in the non diagonal case
\cite{yp}. The constituent quark model relations are summarized in 
Table \ref{nondiag}, which shows that
K-multiplet to k-multiplet and k-multiplet to K-multiplet 
transitions, among {\it P} wave states, are determined by 
four independent couplings $h^{(0)}_p$, $h^{(1)}_p$, $h^{(2)}_p$ and 
$h_f$. We are in agreement with the results of \cite{yp} 
on the $L=0 {\rm and } 2 $ p-wave transitions. Our results on the
$f$-wave transitions 
are new. The contributions from $L=1$ tensor operator 
$P^{(2)}_{\sigma\lambda_1\lambda_2} $ were neglected in \cite{yp} assuming
 that the light quark momenta are not changed in the pion emission process.  
\vspace{5mm}
\begin{table}
 \caption{\label{nondiag} Constituent quark model predictions for non diagonal 
 {\it P} wave to {\it P} wave one pion transitions. } 
 \vspace{3mm}
 \renewcommand{\baselinestretch}{1.2}
 \small \normalsize
 \begin{center}
 \begin{tabular}{|c||c|c|c|c|}
 \hline \hline
  & $h_p^{(0)}$ & $h_p^{(1)}$ & $h_p^{(2)}$ & $h_f$ \\
 \hline \hline
 $ h_{1p}$ & $ 0 $ & $ 0 $ & $ 0 $ & $-$ \\
 \hline
 $ h_{2p}$ & $\frac{1}{\sqrt{3}} $ & $\frac{1}{\sqrt{3}}$ & 
  $\frac{5}{3}\frac{1}{\sqrt{3}} $ & $-$ \\
 \hline
 $ h_{3p}$ & $-\frac{1}{\sqrt{2}} $ & $-\frac{1}{2\sqrt{2}} $ & 
 $\frac{5}{6} \frac{1}{\sqrt{2}} $ & $-$ \\
 \hline
 $ h_{4p,4f}$ & $1 $ & $-\frac{1}{2} $ & 
 $\frac{1}{6} $ & $1$\\
 \hline
 $ h_{5p}$ & $\frac{1}{\sqrt{3}} $ & $-\frac{1}{\sqrt{3}}$ & 
  $\frac{5}{3}\frac{1}{\sqrt{3}} $ & $-$ \\
 \hline
 $ h_{6p}$ & $\sqrt{\frac{2}{3}} $ & $-\frac{1}{\sqrt{6}}$ & 
  $-\frac{5}{3}\frac{1}{\sqrt{6}} $ & $-$ \\
 \hline
 $ h_{7p}$ & $-\frac{1}{\sqrt{2}} $ & $\frac{1}{2\sqrt{2}} $ & 
 $\frac{5}{6} \frac{1}{\sqrt{2}} $ & $-$ \\
 \hline
 $ h_{8p}$ & $\sqrt{\frac{2}{3}} $ & $\frac{1}{\sqrt{6}}$ & 
  $-\frac{5}{3}\frac{1}{\sqrt{6}} $ & $-$ \\
 \hline
 $ h_{9p}$ & $-\frac{1}{2} $ & $ 0 $ & 
  $-\frac{5}{6} $ & $-$ \\
 \hline
 $ h_{10p,10f}$ & $\frac{1}{\sqrt{2}} $ & $-\frac{1}{\sqrt{2}} $ & 
 $ \frac{\sqrt{2}}{3} $ & $-\frac{1}{\sqrt{2}}$\\ 
 \hline
 $ h_{11p,11f}$ & $ 1 $ & $ \frac{1}{2}$ & 
  $\frac{1}{6} $ & $1$ \\
 \hline
 $ h_{12p,12f}$ & $-\frac{1}{\sqrt{2}} $ & $-\frac{1}{\sqrt{2}} $ & 
 $-\frac{\sqrt{2}}{3} $ & $-\frac{1}{\sqrt{2}}$\\ 
 \hline
 $ h_{13p,13f}$ & $ 1 $ & $ 0 $ & 
  $-\frac{1}{3} $ & $1$ \\
 \hline \hline
\end{tabular}
 \renewcommand{\baselinestretch}{1}
 \small \normalsize
 \end{center}
 \end{table} 
 \newpage 
\newpage
 \setcounter{equation}{0}
\def\theequation{B.\arabic{equation}}
\section*{Appendix B \\ Recoupling coefficient approach to diquark 
transitions in the constituent quark model} 
 In this appendix, we shall make use of the quantum theory of angular momentum 
 to describe one-pion transitions between heavy 
 baryon states in the constituent quark model. We show that all
 one-pion transition matrix elements 
 can be written in a very compact form in terms of a product of a 6j-
 and 9j-symbols 
 and corresponding reduced matrix elements. This derivation and the results
 are completely
 equivalent to the covariant coupling approach used in the main text.
 We have added the material in this Appendix for those of our readers
 who are more familiar with the recoupling approach to angular momentum
 transitions in composite systems than the covariant approach used in
 the previous sections. We should mention that we will omit flavour
 factors in our discussion of the one-pion transitions in this appendix.  
 These factors can be obtained from Eq. (\ref{I-factors}) as described at
the
end of the second paragraph in Appendix A.    
  
  As we have discussed before, in the heavy quark limit the pion is
 coupled to the light diquark system and the heavy quark does not
 participate in the transition process. The number of angular momenta involved 
  in the single pion transition between the two light diquark systems
 total $12$ altogether. 
    These angular momenta belong to three different angular
 momentum spaces, the 
  spin, the orbital angular momentum and the total angular momentum spaces.
  In the spin 
  space, we have the spectator quark spin $S_{q_s}=\frac{1}{2}$, the initial 
  and final active quark spins $S_{q_1}=\frac{1}{2}$ and $S_{q_2}=\frac{1}{2}$, 
   respectively, and the quark level one-pion transition 
  operator ${\cal O}^{\sigma}$ with spin $1$ assuming that the one-pion
 transition is due to one-body interaction. Also, we have 
  the initial and 
  final diquark spins $S_1=0,1$ and $S_2=0,1$, respectively. In the orbital
  angular momentum 
  space, one has 
  the diquark initial and final orbital angular momenta $L_1$ and $L_2$ 
  as well as the orbital transition operator ${\cal O}^L$ with
  $ \mid L_1-L_2\mid \leq L \leq \mid
  L_1+L_2 \mid $. And, finally, there are the initial and final diquark total 
  angular momenta $j_1$ and $j_2$, respectively, and the diquark pion 
  emission operator ${\cal O}^l$ with angular momentum $l$ which is even or
  odd depending on the parity of 
  the diquark state. These last three 
  angular momenta operate in the total angular momentum space. 
  
   Since there are $12$ angular momenta involved in the single pion transition, 
    one would presume, at first sight, that these transitions 
   are described in terms of a 12-j symbol. One notices that  
   the spin and orbital spaces, however, factories
   when the spin-orbit coupling is neglected. Hence, as we shall demonstrate later on, the one-pion
   transitions can be written in terms of a 
   product of a 6j- and 9j-symbols and the corresponding reduced
   matrix elements. 
   
   First, let us write down the relevant two coupling schemes in spin 
   space and in orbital angular momentum space as well as the
   relevant recoupling 
   coefficients between the two respective coupling schemes. In the spin 
   space, we have \\ 
   {\bf Coupling Scheme I:} 
   \be 
    \vec{ S}_{q_s}+\vec{ S}_{q_2}=
   \vec{ S}_2  \;\;,\;\;  \vec{ S}_2+\vec{ \sigma}=\vec{ S}_1   
   \ee
   {\bf Coupling Scheme II:}
   \be
    \vec{ S}_{q_2}+\vec{ \sigma}= 
   \vec{ S}_{q_1}  \;\;,\;\;  \vec{ S}_{q_s}+\vec{ S}_{q_1}=\vec{ S}_1 \; ,
  \ee
 with the recoupling coefficient (6j-symbol) $
  \left\{\begin{array}{rrr}
     S_{q_s} & S_{q_2} & S_2\\
      \sigma & S_1 & S_{q_1}  
  \end{array} \right \} $ . 
       
  In  orbital angular momentum space one has\\
   {\bf  Coupling Scheme I:} 
   \be
   \vec{ L}+\vec{ \sigma}=\vec{ l} \;\;,\;\;
    \vec{ L}_2+\vec{ S}_2=\vec{ j}_2  \;\;,\;\; \vec{ l}+\vec{ j}_2=
    \vec{ j}_1 
    \ee
   {\bf Coupling Scheme II:}  
   \be
   \vec{ L }+\vec{ L}_2=\vec{ L }_1 
   \;\;,\;\;  \vec{ \sigma}+\vec{ S}_2=\vec{ S}_1  \;\;,\;\; 
    \vec{ L}_1+\vec{ S}_1=\vec{ j}_1 \; , 
    \ee 
  and the recoupling coefficient (9j-symbol) is given by 
  $ \left\{
  \begin{array}{rrr}
     L & \sigma & l \\
     L_2 & S_2 & j_2 \\
      L_1 & S_1 & j_1 
  \end{array} 
  \right\} $ .
A pictorial representation of the 6j- and 9j- symbols both in orbital and
spin spaces are shown in Figure 1. In these diagrams, the links
represent the orbital (spin) angular momenta while the nodes represent
their couplings.   
  
  Next, one needs to define reduced matrix elements for the transition 
  amplitudes in spin, orbital angular momentum and total angular momentum
  space. 
  We use the Wigner-Eckart theorem 
  and the conventions of Ref. \cite{vmk} to write
  \begin{eqnarray}
  \langle S_{q_2}m_{q_2} \mid {\cal O}^{\sigma}(m_{\sigma}) 
  \mid S_{q_1}m_{q_1} \rangle &=&C^{S_{q_2}m_{q_2}}_{S_{q_1}m_{q_1}
  \sigma m_{\sigma}}\frac{\langle S_{q_2} \mid\mid {\cal O}^{\sigma} 
  \mid \mid S_{q_1} \rangle }{\sqrt{2S_{q_2}+1}} \\
  \langle L_2 m_{L_2} \mid {\cal O}^{L}(m_L) 
  \mid L_1 m_{L_1} \rangle &=&C^{L_2 m_{L_2}}_{{L_1}m_{L_1}
  L m_L}\frac{\langle L_2 \mid\mid {\cal O}^{L} 
  \mid \mid L_1 \rangle }{\sqrt{2L_2+1}} \\
  \langle j_2 m_{j_2} \mid {\cal O}^{l}(m_l) 
  \mid j_1 m_{j_1} \rangle &=&C^{j_2 m_{j_2}}_{{j_1}m_{j_1}
  l m_l}\frac{\langle j_2 \mid\mid {\cal O}^{l} 
  \mid \mid j_1 \rangle }{\sqrt{2j_2+1}} \;, 
  \end{eqnarray}
  here, $C^{im_i}_{jm_jkm_k}$ are Clebsch-Gordan Coefficients (C. G.).
 The total diquark pion emission operator ${\cal O}^{l}(m_{l})$ 
 can be written as a product of the quark level one-pion transition 
 operator ${\cal O}^{\sigma}(m_{\sigma})$ and the orbital transition operator 
 ${\cal O}^{L}(m_L)$ according to
 \be
 {\cal O}^{l}(m_l)=\sum_{m_{\sigma},m_L}C^{lm_l}_{\sigma_{m_{\sigma}}Lm_L}
 {\cal O}^{\sigma}(m_{\sigma}){\cal O}^{L}(m_L) \;.
 \ee
 One can then relate the reduced matrix element of the diquark transition 
 to the product of the reduced matrix elements of the quark level transition
 and the orbital transition. This can be done using the relevant identities 
 for 6j- and 9j- symbols \cite{vmk}. After a little bit of algebra, one
obtains
 \[
 \langle j_2 \mid\mid {\cal O}^{l} 
  \mid \mid j_1 \rangle =(-1)^{S_{q_s}+S_{q_1}+l+S_1+j_1-j_2} \]
  \[
  \sqrt{(2l+1)(2j_1+1)(2j_2+1)(2S_1+1)(2S_2+1)} \]
 \be\label{momenta} 
 \left\{\begin{array}{rrr}
     S_{q_s} & S_{q_1} & S_1\\
      \sigma & S_2 & S_{q_2}  
  \end{array} \right \} 
  \left\{
  \begin{array}{rrr}
     \sigma & L & l \\
     S_1 & L_1 & j_1 \\
      S_2 & L_2 & j_2 
  \end{array} 
  \right\} \langle S_{q_2} \mid\mid {\cal O}^{\sigma} \mid \mid S_{q_1} \rangle 
  \langle L_2 \mid\mid {\cal O}^{L} \mid \mid L_1 \rangle \;. 
 \ee
  Up to a proportionality and phase space factors, the reduced diquark matrix 
  elements of the diquark transition 
  $\langle j_2 \mid\mid {\cal O}^{l} \mid \mid j_1 \rangle $   
  corresponds to the HQS coupling factors $f_{l_{\pi}}$ 
  defined in Eq. (\ref{trans}) and in Table \ref{tabhqs}. 
On the other hand, the 
  product of the reduced matrix elements 
  $\langle S_{q_2} \mid\mid {\cal O}^{\sigma} \mid \mid S_{q_1} \rangle $ and
  $\langle L_2 \mid\mid {\cal O}^{l} \mid \mid L_1 \rangle $
  corresponds to the coupling factors $f_p$, $f_s^{(K,k)}$, 
  $f_d^{(K,k)}$ etc. defined in Sec. 3. 
  The relations between the reduced matrix elements 
  Eq. (\ref{momenta}) correspond to the relations given in Eqs. 
  (\ref{fp}-\ref{fk}) for {\it S} wave to {\it S} wave and {\it P} wave to
{\it S} wave
transitions
  and the relations given in Appendix (A) for {\it P} wave to {\it P} wave
single pion 
  decays.  
  
  For the sake of completeness we shall also give the relation between the 
  reduced matrix elements of the heavy baryon transitions and the light 
  diquark transitions \cite{Z93,fp93}. Using the conventions of 
  \cite{vmk}, the reduced matrix elements for the one-pion transition between 
  heavy baryons is defined by  
  \be
  \langle J_2 M_2 \mid {\cal O}^{l}(m_l) 
  \mid J_1 M_1 \rangle =C^{J_2 M_2}_{{J_1}M_1
  l m_l}\frac{\langle J_2 \mid\mid {\cal O}^{l} 
  \mid \mid J_1 \rangle }{\sqrt{2J_2+1}} \; .
  \ee
  In the heavy quark limit one has
  \[
 \langle J_2 \mid\mid {\cal O}^{l} 
  \mid \mid J_1 \rangle =(-1)^{J_1+\frac{1}{2}+l-j_2} 
  \sqrt{(2J_1+1)(2J_2+1)} \] \be
 \left\{\begin{array}{rrr}
     l & j_1& j_2\\
      \frac{1}{2} & J_2 & J_1  
  \end{array} \right \} 
  \langle j_2 \mid\mid {\cal O}^{l} \mid \mid j_1 \rangle \;. 
  \ee   
The pictorial representation of the 6j-symbol appearing in this equation
is shown in Figure 2.  
To close this appendix, we also give values \cite{vmk} for 6j-symbols 
  required to calculate strong decay rates of Eq. (\ref{rate-gen}) for all 
  transitions among heavy baryon states.  
  \begin{table}
 \caption{\label{tab6j}
 Table of values for 6j-symbols
 necessary to calculate the decay rates in Eq.(\ref{rate-gen}) where 
$s=l+j_1+j_2$.}
 \vspace{5mm}
 \renewcommand{\baselinestretch}{1.2}
\small \normalsize
  \begin{center}
  \begin{tabular}{|c||c|c|}
   \hline \hline
    & $J_2=j_2+\frac{1}{2}$ & $J_2=j_2-\frac{1}{2}$   \\
   \hline \hline
   $J_1=j_1+\frac{1}{2}$  &
    $(-1)^{(s+1)}\frac{1}{2}\left[ \frac{(s+2)(s-2l+1)}{(2j_1+1)(j_1+1)
 (2j_2+1)(j_2+1)} \right]^{1/2} $ &
  $(-1)^{s}\frac{1}{2}\left[ \frac{(s-2j_2+1)(s-2j_1)}{(2j_1+1)(j_1+1)
  j_2(2j_2+1)} \right]^{1/2} $ \\
  \hline
 $J_1=j_1-\frac{1}{2} $ &
 $(-1)^{s}\frac{1}{2}\left[ \frac{(s-2j_2)(s-2j_1+1)}{j_1(2j_1+1)
  (2j_2+1)(j_2+1)} \right]^{1/2} $ &
 $(-1)^{s}\frac{1}{2}\left[ \frac{(s+1)(s-2l)}{j_1(2j_1+1)
  j_2(2j_2+1)} \right]^{1/2} $ \\
 \hline \hline
 \end{tabular}
 \renewcommand{\baselinestretch}{1}
 \small \normalsize
 \end{center}
 \end{table}
 \newpage
\section*{Acknowledgments}
  We would like to acknowledge informative discussions with D. Pirjol.
  We would, also,
  like to thank A. Ilakovac, U. Kilian and J. Landgraf for their
  participation in the
  early stages of this work. S. T. would like to thank 
  Patrick O. J. O'Donnell and the Department of Physics, University of 
  Toronto, Toronto, Canada for the hospitality during the final stages of 
  this work and also for reading the manuscript. J.G.K. acknowledges partial 
  support from the BMBF (Germany) under contract 06MZ865.

\newpage 
\begin{center}
  {\large \bf Figure Captions}
\end{center}
\vspace*{2cm}
Fig. 1: Constituent quark model recoupling diagrams representing a 6j- and
9j-symbols a) 6j-symbol acting in spin space b) 9j-symbol acting in
orbital angular momentum space. Links represent angular momenta
and nodes represent their couplings. The Angular momenta are defined in
the text \\ \\
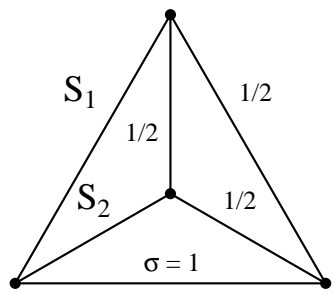
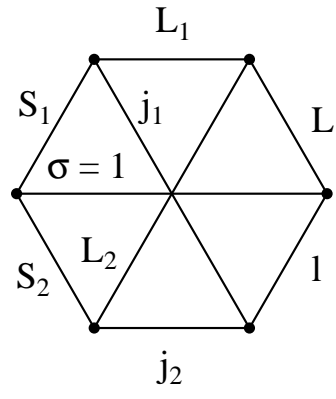
Fig. 1: The HQS limit recoupling diagram representing a 6j-symbol acting
in total angular momentum space. Links represent angular momenta
and nodes represent their couplings. The Angular momenta are defined in
the text \\ \\ 
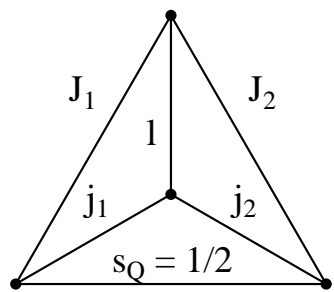
\end{document}